\documentstyle[11pt,newpasp,twoside]{article}
\begin{document}
\title{The Bologna Key Project on $\omega$~Centauri}
\author{Francesco R. Ferraro}
 \affil{Osservatorio Astonomico, Via Ranzani 1, I-40127
  Bologna, Italy}
\author{Elena Pancino}
 \affil{Dipartimento di Astronomia, Universit\`a di Bologna, Via
  Ranzani 1, I-40127 Bologna, Italy}
 \affil{European Southern Observatory, K. Schwarzschildstr. 2,
  D-85748 Garching bei M\"unchen, Germany}
\author{Michele Bellazzini}
 \affil{Osservatorio Astonomico, Via Ranzani 1, I-40127
  Bologna, Italy}


\begin{abstract}
In this contribution we schematically summarize a number of ongoing,
coordinated spectro-photometric projects devoted to the study of the
stellar population properties in $\omega$~Cen. All of these
investigations are part of a global comprehensive project led by our
team to address the complex formation history of stars in this
cluster.
\end{abstract}


\section{Introduction}

After the recent discovery of a peculiar, extremely metal rich RGB
population, we started a long term project devoted to the study of the
multiple stellar populations in $\omega$~Cen. The global project
consists of two main branches: {\it (a)} complete multi-wavelength
photometric surveys to investigate the general properties of the
sub-populations, from the far ultraviolet to the near infrared, using
both wide-field and high-resolution imaging; {\it (b)} high and
moderate resolution, high signal-to noise, extensive spectroscopic
campaigns for giant and subgiant stars both in the optical and the
infrared wavelength ranges. As a final result, we expect to completely
characterize the abundance patterns, the dynamical properties and the
relative ages of the sub-populations in $\omega$ Cen. A short
description of some ongoing sub-projects is provided, along with some
first preliminary results.

\section{The Wide Field Photometric Campaign}

The large field photometric campaign was performed in 3 different
observing runs during 1999 and 2000 at the 2.2m ESO/MPI telescope (La
Silla, Chile), equipped with the Wide Field Imager (WFI). We covered a
large area (33$^{\prime} \times$33$^{\prime}$) around the cluster and
more than 230,000 stars were measured. The $(B,B-I)$ color magnitude
diagram (CMD) based on this dataset led to the identification of an
anomalous Red Giant Branch (RGB) sequence (hereafter RGB-a)
significantly redder and probably more metal rich than the bulk
population in $\omega$~Cen (Pancino et al. 2000; Pancino, this
volume). During the same observing runs we have also obtained
$U$,$V$,$H_{\alpha}$ and $R$ images which are now under analysis.

\begin{figure}
\plotfiddle{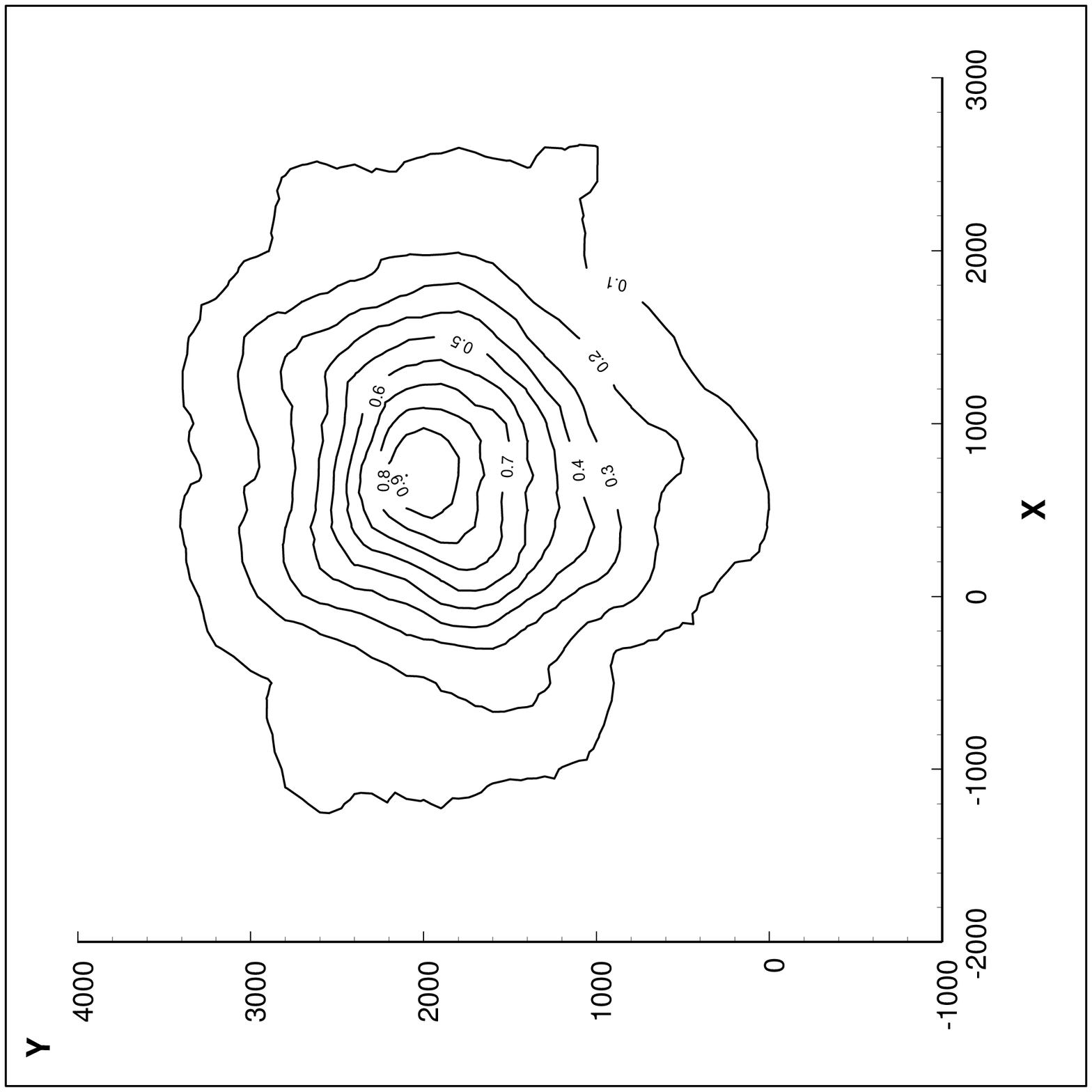}{6cm}{270}{33}{33}{-190}{175}
\plotfiddle{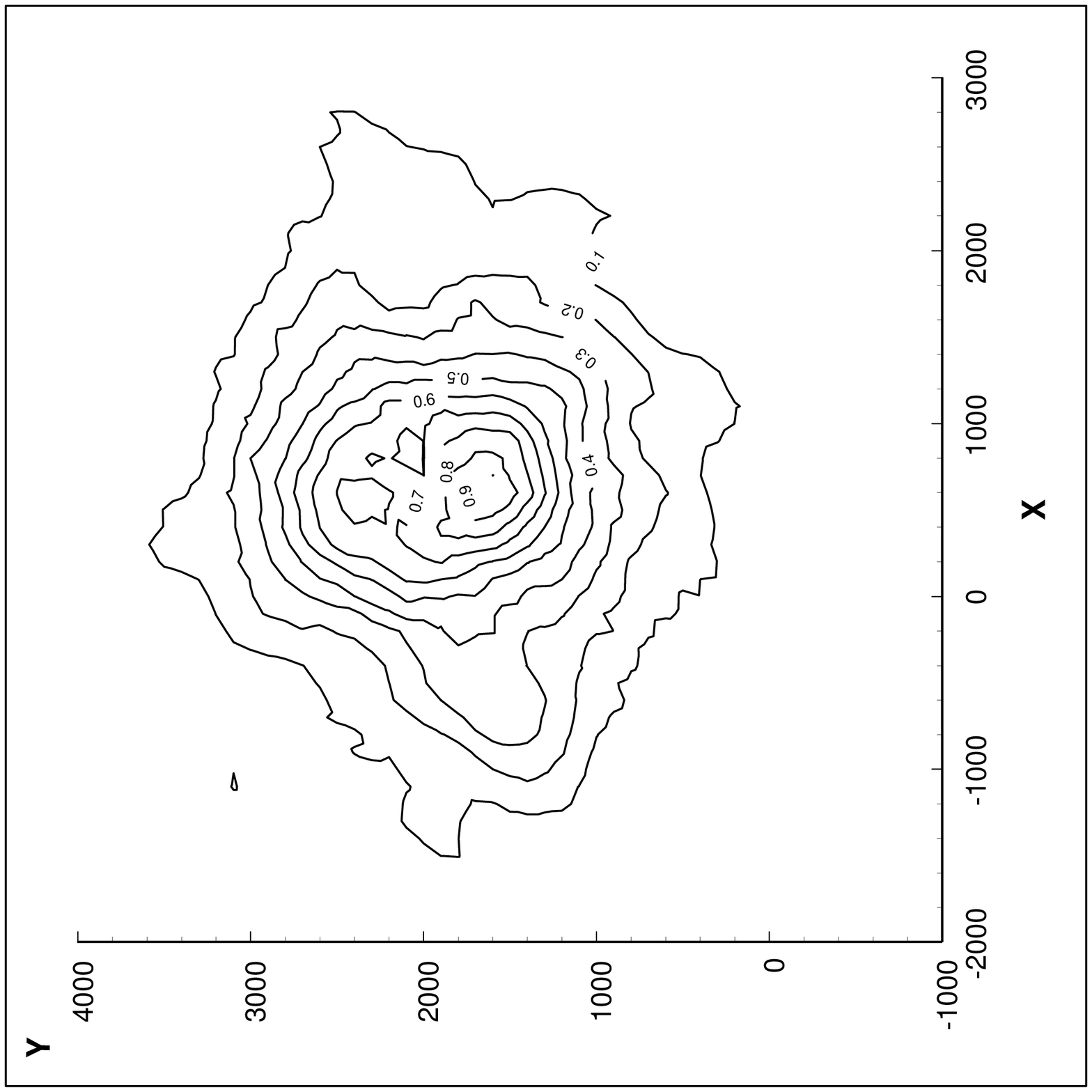}{0cm}{270}{33}{33}{10}{199}
\caption{The spatial distribution for the RGB-MP ({\it Left 
Panel}) and RGB-MInt ({\it Right Panel}) populations are compared. The
different elongations are well visible.}
\end{figure}

Following Pancino et al. (2000), three different sub-populations of
red giants with different metallicities can be defined: {\it (a)} a
metal poor population with [Ca/H]$\sim-1.4$ (RGB-MP); {\it (b)} a
metal intermediate one with $-1.1<$[Ca/H]$<-0.5$ (RGB-MInt) and {\it
(c)} a metal rich, anomalous one with [Ca/H]$\sim -0.1$ (RGB-a).

The spatial distributions for the RGB-MP and RGB-MInt populations are
shown in Figure~1. As can be seen, the RGB-MP population shows the
well known East-West elongation for $\omega$~Cen. The RGB-MINT one is
instead elongated along the North-South direction, and shows
complicated isopleths. Generalized (bidimensional) Kolmogorov-Smirnov
tests ensure that the distributions of the two sub-samples are not
compatible, suggesting that the two sub-populations have different
dynamical properties.

\subsection{The Tip of the Red Giant Branch}

\begin{figure}
\plotfiddle{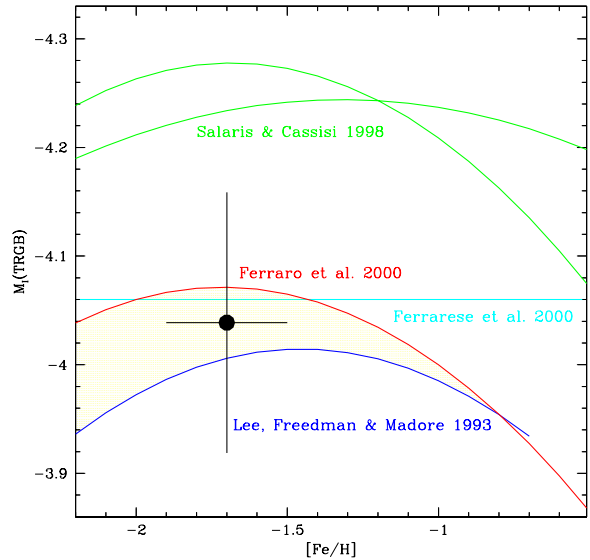}{6.5cm}{0}{40}{40}{-130}{-70}
\caption{Our calibrating point for $\omega$~Cen (black dot), based on
the recent detached eclipsing binary distance esimate by Thompson et
al (2001) and on the photometry by Pancino et al. (2000), is compared
with different calibrations for the RGB Tip method.}
\end{figure}

In a recent paper by Bellazzini, Ferraro \& Pancino (2001), the
photometric dataset by Pancino et al. (2000) was used in order to
derive the RGB-Tip magnitude in $\omega$~Cen. The large number of
giants measured in this cluster (more than 1700 in the brightest 3
mags) allows us to apply the same technique adopted by the HST key
project on the extragalactic distance scale, obtaining
$I^{TRGB}=9.84\pm 0.04$. Using the distance by Thompson et al. (2001)
we obtained $M_I^{TRGB}=-4.04\pm 0.12$. {\it This value represents the
most accurate empirical anchor point for the calibration of the
$M_I^{TRGB}$--[Fe/H] relation at [Fe/H]=--1.7.}

In Figure~2 this measure is plotted as a filled dot and compared with
other empirical and theoretical calibrations. As can be seen, our
value lies just in the region where most of the empirical relations
lie (shaded area).

\begin{figure}
\plotfiddle{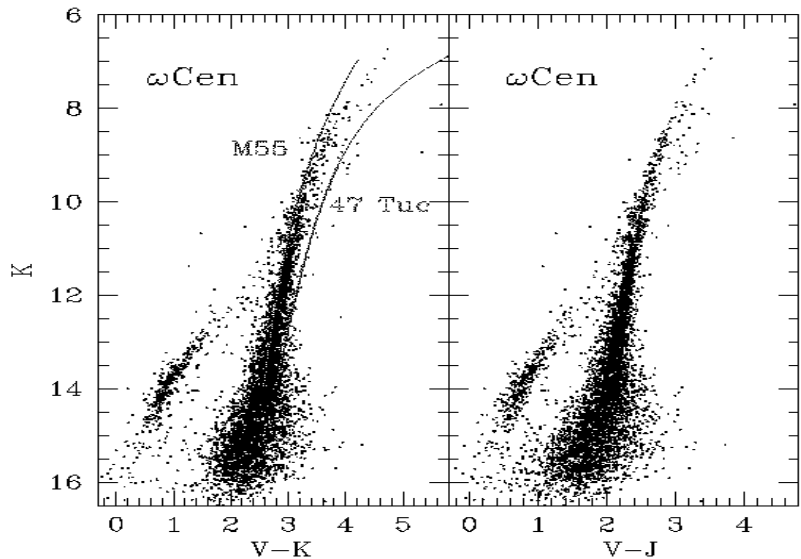}{7.5cm}{0}{55}{40}{-180}{-70}
\caption{Preliminary results from the SOFI infrared photometry: the
IR-optical CMD for the central region of the cluster is shown. The
mean RGB ridge lines for M~55 and 47~Tuc from Ferraro et al. (2000)
are overplotted for reference.}
\end{figure}

\section{Infrared Photometry}

A mosaic of 3$\times$3 images in $J$ and $K$ was secured during an
observing run at NTT (La Silla), with SOFI. The total field covered is
roughly $\sim13^{\prime}\times 13^{\prime}$ around the cluster
center. Infrared photometry is especially important in deriving
structural parameters of cool giant stars. In particular, combined
optical-infrared colours such as $(V-K)$ are powerful photometric
indicators of $T_{eff}$. A preliminary reduction of the central field
(5$^{\prime}\times$5$^{\prime}$) has been already completed. Figure~3
shows the CMD for this region, obtained by combining the $J$ and $K$
with the $V$ magnitude from the wide field observations. The mean RGB
ridge lines of M~55 ([Fe/H]$\sim-$1.81) and 47~Tuc ([Fe/H]$\sim-$0.75)
from Ferraro et al. (2000) are overplotted for reference. As can be
seen from the figure, the existence of the RGB-a is confirmed also by
this diagram: a few stars with metallicity similar to that of 47~Tuc
are present to the right side of the main RGB.

\section{The Spectroscopic Campaign}

We are performing extensive spectroscopic campaigns of stars in
$\omega$~Cen, with the following goals:\\ {\it (i)} To investigate the
detailed chemical properties of the newly discovered ultra metal rich
RGB-a.\\ {\it (ii)} To measure [Fe/H] and other iron peak elements
abundances, and to assess membership of the RGB-a candidates.\\ {\it
(iii)} To identify the major chemical contributors (SNeII, SNeIa, AGB
stars) for the different sub-populations, by measuring $\alpha$, $r$
and $s$-elements abundances.\\ {\it (iv)} To look for CN-CH
anticorrelations and to explore the effects of mixing, of the
environment and of pollution from massive AGB stars.

\begin{figure}
\plotfiddle{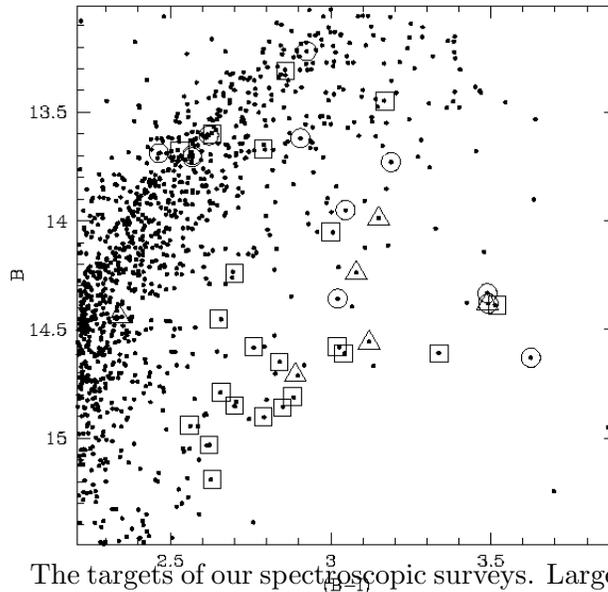}{7.5cm}{0}{40}{40}{-130}{-70}
\caption{The targets of our spectroscopic surveys. Large open
triangles refer to the 6 stars observed during the UVES pilot survey
in june 2000. Large open squares refer to the 23 stars observed with
UVES in apr--may 2001. Large open circles mark the 13 stars observed
with SOFI in the IR.}
\end{figure}

\vspace{0.2cm} 
In the following we schematically summarize the status of the ongoing
spectroscopic surveys:

\vspace{0.2cm} 
\noindent {\it (1)}~High-resolution optical and UV spectroscopy of RGB and
SGB-TO stars using UVES (UV-Visual Echelle Spectrograph) at the ESO
VLT.
\begin{itemize} 
\item{RGB sample: a sample of 29 stars, belonging to the extremely
metal rich, but also to the intermediate and the metal pooor RGB, has
been already observed (see Figure~4). Preliminary results are
presented in the contribution by Pancino, this volume.}
\item{SGB-TO sample: we have been awarded 3 nights of observing time
with UVES in march 2002, to study $\sim10$ stars at the TO level.}
\end{itemize}

\begin{figure}
\plotfiddle{Ferraro_fig05.aph.s}{13.5cm}{0}{65}{70}{-195}{-85}
\caption{Four samples of the IR spectra for three RGB-a program stars are
shown. Stars ST~21 and ST~51 correspond to stars ROA~513 and ROA~300
in the Woolley (1966) catalog. Some of the most prominent atomic and
molecular features are marked.}
\end{figure}

\noindent {\it (2)}~Infrared low and medium resolution campaign. The
near IR spectral range is particularly suitable to study cool giants,
due to the intrinsic sensitivity to low temperatures, and the
background contamination by main sequence stars is much less severe
than in the optical range, allowing to properly characterize the red
stellar sequences also in crowded and/or metal rich environments.  The
near IR ($1-2.5\mu m$) spectra of cool stars show many absorption
lines due to neutral metals and molecules. A reasonable number of
these lines are strong, not heavily saturated and not affected by
severe blending, hence they can also be measured at relatively low
resolution and safely modeled under Local Thermodynamical Equilibrium
(LTE) approximation. Thus, accurate abundances of key elements like
iron, carbon (from the CO molecular bands), oxygen (from the OH
molecular bands) and other $\alpha$-elements can be easily obtained.

We already observed a sample of 13 giants at the NTT (La Silla, Chile)
using SOFI (see Figure~4).  Figure~5 shows four portions of the IR
spectra (with a resolution of 7 and 14\AA) for three RGB-a stars.
Preliminary iron abundance determinations confirm that the RGB-a has a
metallicity higher than [Fe/H]$\sim -0.8/-0.9$. In particular, the
iron abundance for star ROA~300 nicely agrees with the results
obtained with UVES.

\begin{figure}
\plotfiddle{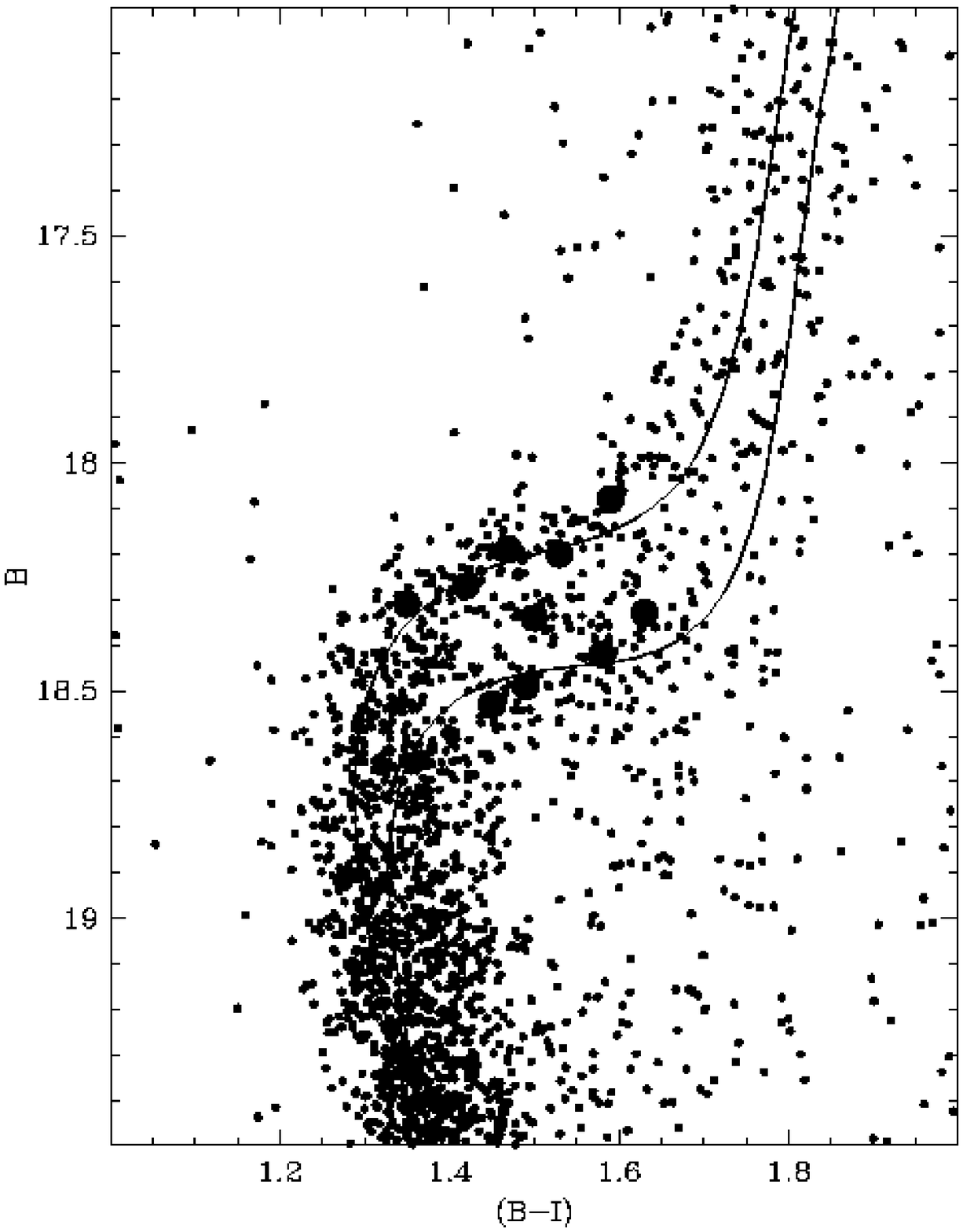}{7.5cm}{0}{33}{33}{-110}{-20}
\caption{The SGB-TO region in the outer parts of $\omega$~Cen from the 
WFI dataset by Pancino et al. (2000). Targets for UVES high-resolution
spectroscopy are plotted as large filled circles. The two lines
represent isochrones with the same age (16$~Gyr$) and different
metallicities ([Fe/H]=$-1.7$ and $-1.3$).}
\end{figure}

\section{The Relative Ages Problem}

A fundamental step in order to understand the star formation history
in this cluster is a precise measure of the ages differences among the
different populations.

Thus, we are also performing a detailed study to investigate weather
an age spread does exist among the various sub-populations in $\omega$
Cen. In order to measure {\em relative ages} we are collecting: {\em
(a)} high resolution UVES spectroscopy of a sample of Sub-Giant Branch
(SGB) stars in the outer region of the cluster, where two distinct SGB
sub-branches can clearly be seen (Figure~6), and {\em (b)} high
precision FORS1 VLT photometry of the central regions of the cluster,
that in our previous WFI photometry are hidden by crowding. The entire
data-set will allow us to disentangle age and metallicity, finally
establishing {\em relative ages} of the sub-populations.

\section{The Future Surveys with FLAMES}

The Bologna Observatory is member of the ITAL-FLAMES consortium (a
consortium of four italian observatories: Bologna, Trieste, Palermo
and Catania) which participated to the construction of the FLAMES
facility. The Fibre Large Array Multi Element Spectrograph (FLAMES) is
a system that feeds with fibre links two spectrographs simultaneously,
GIRAFFE and UVES, located on the Nasmyth platforms of VLT-UT2. GIRAFFE
is a multi-object echelle spectrograph covering the entire visible
range (370-900 nm) at intermediate (R=7500-12500) and high
(R=15000-25000) resolution. FLAMES will therefore allow
intermediate-high resolution spectroscopy for objects brighter than
V=22 with a multiplex capability of 8 to 130 fibers over a large field
of view (about 25 arcmin in diameter).

The use of this facility will allow us to make a significant step
towards the final characterization of the different sub-populations in
$\omega$~Cen, both from the dinamycal and chemical point of view. In
Figure~7 we show a preliminary selection of candidates for the RGB
({\it Left Panel}) and the SGB-TO ({\it Right Panel}) FLAMES surveys.

We plan to use part of the Consortium guaranteed time to perform
simultaneous observations with UVES and GIRAFFE. By mantaining an
overlap of 5-10 stars between the high resolution ($R\sim 40000$) and
the medium resolution ($R\sim 20000$) samples, we can check the
performance of the spectral synthesis method used and avoid using
blended or unreliable lines in the lower resolution spectra. An
important byproduct of this massive spectroscopic survey is the
production of an unprecedented catalog of radial velocity
estimates. We will achieve a precision of 0.15 $km~sec^{-1}$, that
will enable us to explore the correlation between chemical and
dynamical properties in the cluster stars.

\begin{figure}
\plotfiddle{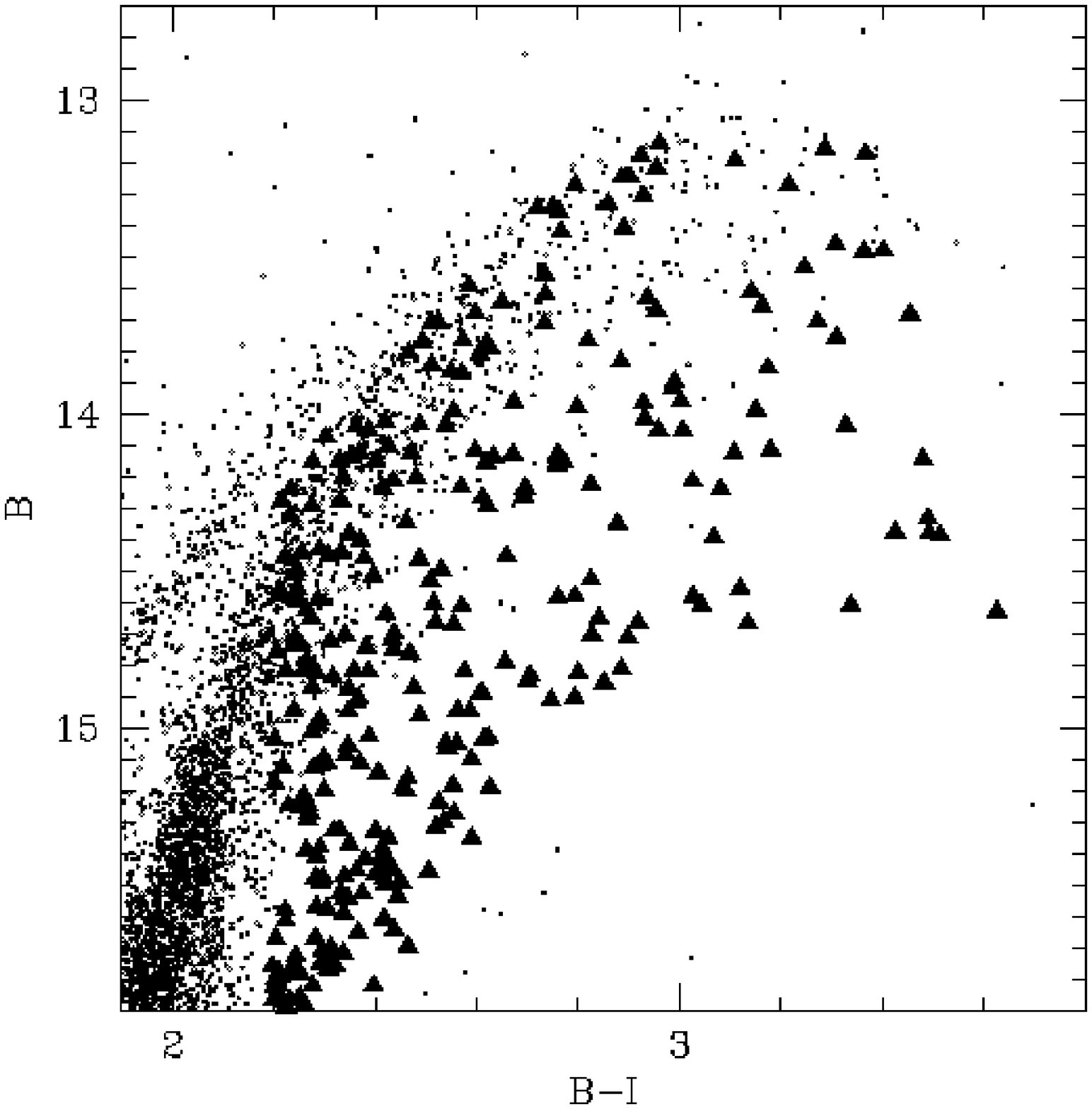}{6.5cm}{0}{35}{40}{-200}{-90}
\plotfiddle{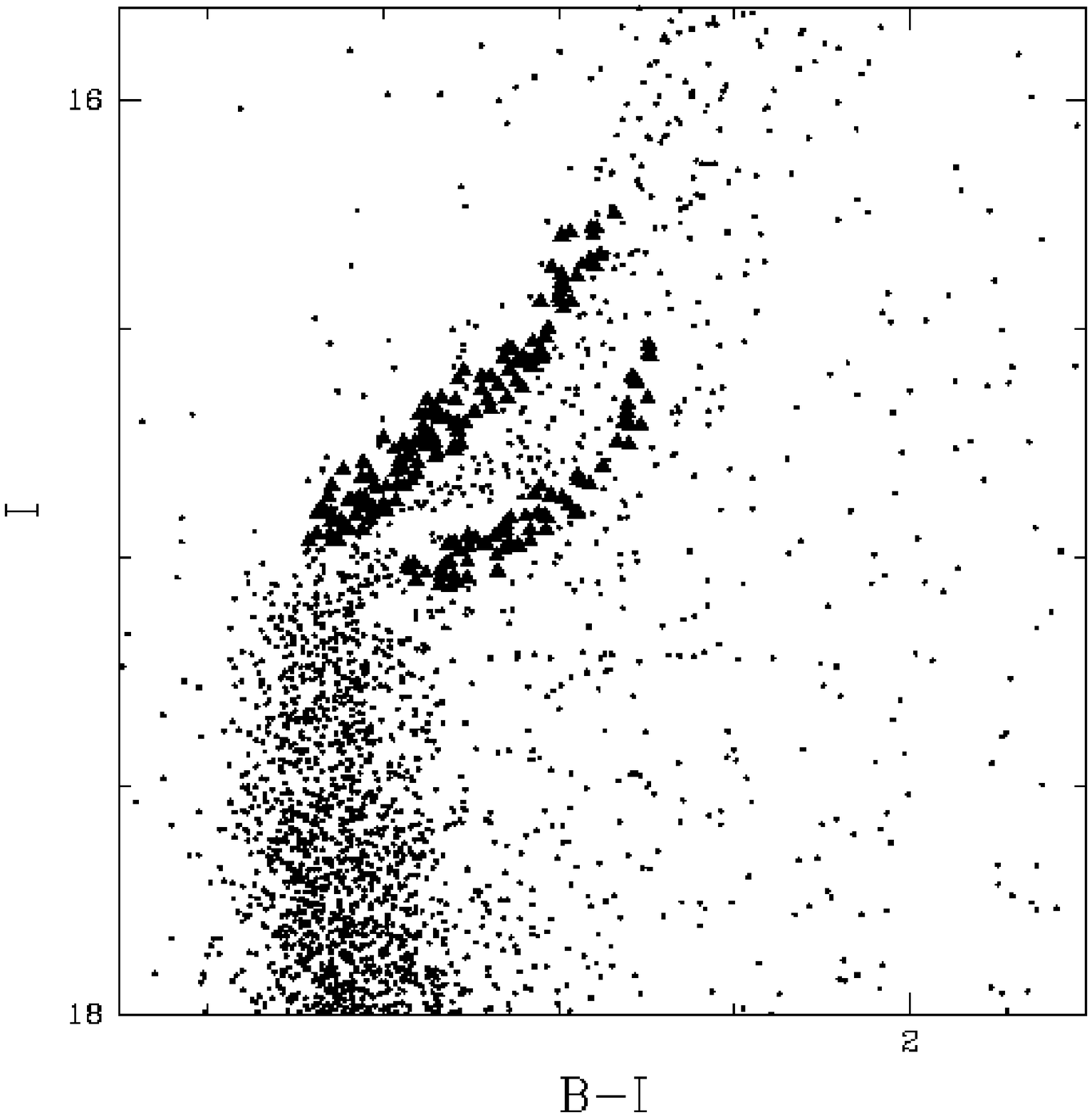}{0cm}{0}{35}{40}{-20}{-65}
\caption{A possible selection of targets for FLAMES surveys. {\it
Left:} the RGB survey. Some stars will be observed both with UVES and
GIRAFFE for comparison. Many of our targets are chosen to be in common
with previous published studies. {\it Right:} the SGB-TO survey.  We
plan to obtain a detailed chemical description of these stars to look
for the counterparts of the RGB sub-populations and finally solve the
relative ages problem.}
\end{figure}


\vspace{0.5cm}
We thank the friends and collaborators that are contributing to this
long term project: A. Seleznev, L. Pasquini, V. Hill, L. Origlia,
L. Monaco, A. Sollima, G. Piotto, E. Pompei, T. Augusteijn.


\end{document}